\documentstyle[12pt,epsf]{article}

\newcommand{\bea}{\begin{eqnarray}}
\newcommand{\eea}{\end{eqnarray}}
\newcommand{\be}{\begin{equation}}
\newcommand{\ee}{\end{equation}}
\newlength{\dinwidth}
\newlength{\dinmargin}
\begin{document}

\centerline{\bf A STRONG ELECTROWEAK SECTOR AT FUTURE LINEAR COLLIDERS$\;^{*)}$}
\vskip 1.0cm
\centerline {R. Casalbuoni$\;^{a,b)}$,
S. De Curtis$\;^{b)}$,
D. Dominici$\;^{a,b)}$,}
\centerline {A. Deandrea$\;^{c)}$, R. Gatto$\;^{c)}$ and
M. Grazzini$\;^{d,e)}$}
\vskip 0.5cm

\noindent
$a)$ Dipartimento di Fisica, Univ. di Firenze, I-50125 Firenze, Italy.
\hfill\break\noindent
$b)$ I.N.F.N., Sezione di Firenze, I-50125 Firenze, Italy.
\hfill\break\noindent
$c)$ D\'ept. de Phys. Th\'eor., Univ. de Gen\`eve, CH-1211 Gen\`eve 4.
\hfill\break\noindent
$d)$ Dipartimento di Fisica, Univ. di Parma, I-43100 Parma, Italy.
\hfill\break\noindent
$e)$ I.N.F.N., Gruppo Collegato Parma,  I-43100 Parma, Italy.
\noindent
\vskip 0.1cm

$^{*)}\;$ Partially supported by the Swiss National Foundation.
This work is part of the EEC project ``Tests
of electroweak symmetry breaking and future european colliders",
CHRXCT94/0579 (OFES 95.0200). Universit\`a di Firenze Preprint-DFF-245/3/1996.
\vskip 0.5cm
\centerline{\bf{ABSTRACT}}
\noindent
An effective lagrangian describing a strong interacting
electroweak sector is considered. It contains new vector and axial-vector
resonances all degenerate in mass and mixed with $W$ and $Z$. 
The model, for large mass of these degenerate gauge bosons, becomes
identical to the standard model in the classical limit of infinite Higgs mass.
The limits on the parameter space of this model from future $e^+e^-$
colliders are presented.

\def\lmu{{{{\bf L}}_\mu}}
\def \LL{{{\bf L}_\mu}}
\def\rmu{{{ \bf R}_\mu}}
\def\rmus{{\cal R}^\mu}
\def\gs{{g''}}
\def\lq{\left [}
\def\rq{\right ]}
\def\dmu{{\partial_\mu}}
\def\dnu{{\partial_\nu}}
\def\dmus{{\partial^\mu}}
\def\dnus{{\partial^\nu}}
\def\gp{g'}
\def\gpt{{{g}^\prime}}
\def\gptd{ g^{\prime 2}}
\def\eps{{\epsilon}}
\def\tr{{ {tr}}}
\def\V{{\cal{V}}}
\def\W{{\bf{W}}}
\def\Wt{\tilde{{W}}}
\def\Y{{\bf{Y}}}
\def\Yt{\tilde{{Y}}}
\def\tW{\tilde W}
\def\tY{\tilde Y} 
\def\tL{\tilde L}
\def\tR{\tilde R}
\def\L{{\cal L}}
\def\s{s_\theta}
\def\c{c_\theta}
\def\gt{\tilde g}
\def\et{\tilde e}
\def\At{\tilde A}
\def\Zt{\tilde Z}
\def\Wpt{\tilde W^+}
\def\Wmt{\tilde W^-}
\def\de{\partial}
\def\eps{\epsilon}
\def\nn{\nonumber}
\def\dd{\displaystyle}
\def\ct{c_\theta}
\def\st{s_\theta}
\def\cdt{c_{2\theta}}
\def\sdt{s_{2\theta}}
\def\qq{<{\overline u}u>}
\def\Lt{{{\tilde L}}}
\def\Rt{{{\tilde R}}}
\def\st{\tilde s_\theta}
\def\ct{\tilde c_\theta}
\def\gt{\tilde g}
\def\et{\tilde e}
\def\A{\bf A}
\def\Z{\bf Z}
\def\Wpt{\tilde W^+}
\def\Wmt{\tilde W^-}

\section{The Model}

In view of future projects of $e^+e^-$ linear colliders it is important
to study the possible phenomenology at such colliders from a strong electroweak
sector \cite{gen}. 
We shall study the effects of the strong electroweak sector af future linear 
colliders, assuming a low energy effective theory. The effective lagrangian 
contains vector and axial-vector resonances as the most 
visible manifestations at low energy of the strong interacting sector 
\cite{dege}.
This model is an extension of the BESS model where only new vector resonances 
are present \cite{bess}. It leads to an interesting and appealing phenomenology.

Let us call $G$ the symmetry group of the theory, spontaneously broken.
Among the Goldstone bosons, three are absorbed to give mass to
$W$ and $Z$. The vector and axial-vector
mesons will transform under the unbroken subgroup $H$ of $G$.
Following the {\it hidden gauge symmetry} approach \cite{bala}\cite{bando},
theories with
non linearly realized symmetry $G\to H$ can be linearly realized by
enlarging the gauge symmetry $G$ to $G\otimes H^\prime\to 
H_D=diag(H\otimes H^\prime)$. $H^\prime$ is a local gauge group 
and the vector and axial-vector are the gauge fields associated to $H^\prime$. 

Let us consider such an effective lagrangian parameterization
for the electroweak symmetry breaking, using $G=SU(2)_L\otimes SU(2)_R$,
$H'=SU(2)_L \otimes SU(2)_R$.
The nine Goldstone bosons resulting from the spontaneous breaking 
of $G'=G\otimes H'$ to $H_D$, can be described by three independent
$SU(2)$ elements: $L$, $R$ and $M$, with the
following transformations properties
\be
L'= g_L L h_L,\quad R'= g_R R h_R,\quad M'= h_R^\dagger M h_L
\label{2.1}
\ee
with $g_{L,R}\in SU(2)_{L,R}\subset G$ and $h_{L,R}\in H'$. 
Moreover we shall require the 
invariance under the discrete left-right transformation 
$P:\quad L\leftrightarrow R,\quad M\leftrightarrow M^\dagger$
which combined with the usual space inversion allows to build the
parity transformation on the fields. If we ignore the transformations of 
eq. (\ref{2.1}), the largest possible global 
symmetry of the low-energy theory is given by the requirement of maintaining 
for the transformed variables $L'$, $R'$ and $M'$ the character of $SU(2)$ 
elements, or $G_{max}=[SU(2)\otimes SU(2)]^3$, consisting of three independent 
$SU(2)\otimes SU(2)$ factors, acting on each of the three variables 
separately. As we shall see, it happens that, for specific choices of the 
parameters of the theory, the symmetry $G'$ gets enlarged to $G_{max}$
\cite{lett}.

The most general $G'\otimes P$ invariant lagrangian is given by \cite{assiali}
\be
{L}_G=-{{v^2}\over {4}} [a_1 I_1 + a_2 I_2 + a_3 I_3 + a_4 I_4]
\ee
plus the kinetic terms ${L}_{kin}$.  The four invariant
terms $I_i$ ($i=1,...4$) are 
given by:
\be
I_1=tr[(V_0-V_1-V_2)^2]\quad
I_2=tr[(V_0+V_2)^2]\quad
I_3=tr[(V_0-V_2)^2]\quad
I_4=tr[V_1^2]
\ee
where
\be
V_0^\mu=L^\dagger D^\mu L\quad
V_1^\mu=M^\dagger D^\mu M\quad
V_2^\mu=M^\dagger(R^\dagger D^\mu R)M
\ee
and the covariant derivatives are
\be
D_\mu L=\partial_\mu L -L \lmu\quad
D_\mu R=\partial_\mu R -R \rmu
\ee
\be
D_\mu M=\partial_\mu M -M \lmu+\rmu M
\ee
where $\lmu (\rmu)$ are gauge fields of $SU(2)_{L(R)}\subset H^\prime$
(instead of working with
vector and axial-vector we work with these left and right combinations).

The kinetic terms are given by
\be
{L}_{kin}={{1}\over{\gs^2}} tr[F_{\mu\nu}({\bf L})]^2+
	 {{1}\over{\gs^2}}  tr[F_{\mu\nu}({\bf R})]^2
\ee
where $\gs$ is the gauge coupling constant for the gauge fields $\lmu$ and
$\rmu$, 
and 
$F_{\mu\nu}({\bf L})$, $F_{\mu\nu}({\bf R})$ are the usual field tensors.

The model we will consider is characterized by the following choice 
of parameters $a_4=0$, $a_2=a_3$ \cite{dege,lett}.
In order to discuss the symmetry properties that make such a choice natural 
it is useful to observe that the invariant $I_1$ could be re-written as 
$
I_1=-tr(\partial_\mu U^\dagger \partial^\mu U)
$
with $U=L M^\dagger R^\dagger$ and the lagrangian as
\be
{L}_G={{v^2}\over {4}}\{a_1~ tr(\partial_\mu U^\dagger \partial^\mu U) +
			 2~a_2~ [tr(D_\mu L^\dagger D^\mu L)+
			  tr(D_\mu R^\dagger D^\mu R)]\}
\label{2.3}
\ee
Each of the three terms in the above expression
is invariant under an independent $SU(2)\otimes SU(2)$ group
\be
U'=\omega_L U \omega_R^\dagger,\quad L'= g_L L h_L,\quad R'= g_R R h_R
\ee
The overall symmetry is $G_{max}=[SU(2)\otimes SU(2)]^3$, with a part 
$H'$ realized as a gauge symmetry.
With the particular choice $a_4=0$, $a_3=a_2$, as we see from eq. (\ref{2.3}),
the mixing between $\lmu$ and $\rmu$ is vanishing, and the new states are 
degenerate in mass.
Moreover, as it follows from eq. (\ref{2.3}), the longitudinal modes
of the fields are entirely
provided by the would-be Goldstone bosons in $L$ and $R$. This means
that the pseudoscalar particles remaining as physical states in the
low-energy spectrum are those associated to $U$. They in turn can
provide the longitudinal components to the $W$ and $Z$ particles,
in an effective description of the electroweak breaking sector.

The peculiar feature of the model is that the new bosons are
not coupled to those Goldstone bosons which are absorbed to give
mass to $W$ and $Z$. As a consequence the channels $W_L Z_L$ and
$W_L W_L$ are not strongly enhanced as it usually happens in models with a
strongly interacting symmetry breaking sector \cite{bess}.
This also implies larger branching ratios of the new resonances into 
fermion pairs.

The coupling of the model to the electroweak 
$SU(2)_W\otimes U(1)_Y$ gauge fields is obtained
via the minimal substitution
\be
D_\mu L \to D_\mu L+ {\W}_\mu L\quad
D_\mu R \to D_\mu R+ {\Y}_\mu R\quad
D_\mu M \to D_\mu M 
\ee
where
\bea
\W_\mu&=&i g {\tilde W}_\mu ^a{{\tau^a}\over {2}}\quad
\Y_\mu=i\gp
 {\tilde Y}_\mu{{\tau^3} \over {2}} \cr
\LL&=&i{{\gs}\over {\sqrt{2}}} {\tilde L}_\mu ^a{{\tau^a}\over {2}}\quad
{\bf R}_\mu=i{{\gs}\over {\sqrt{2}}} {\tilde R}_\mu ^a{{\tau^a}\over{2}}
\eea
with $g$, $\gp$ the $SU(2)_W\otimes U(1)_Y$ 
gauge coupling constant and $\tau^a$ the Pauli matrices.
We have used tilded quantities to reserve untilded variables
for mass eigenstates.

By introducing the canonical kinetic terms for ${\tilde W}_\mu^a$ and 
${\tilde Y}_\mu$ 
and going into the unitary gauge we get
\bea
\L&=&-{{v^2}\over {4}}\Big[ a_1 \tr(\Wt_\mu-\Yt_\mu)^2
+2 a_2 \tr(\Wt_\mu-{\Lt}_\mu)^2
+2 a_2 \tr(\Yt_\mu-\Rt_\mu)^2\Big]\cr
&+&\L^{kin}(\Wt,\Yt,\Lt,\Rt)
\label{2.4}
\eea
The standard model (SM) relations are obtained in the limit $\gs \gg {g},
{g}'$. Actually, for very large $\gs$, the kinetic terms for the fields 
${\tilde L_{\mu}}$ and ${\tilde R_{\mu}}$
drop out, and ${\cal L}$ reduces to the first term in eq. (\ref{2.4}).
This term reproduces precisely the mass term
for the ordinary gauge vector bosons in the SM, provided we 
assume  $a_1=1$.
Finally let us consider the couplings to the fermions:
\newpage
\bea
\L_{fermion} &=& \overline{\psi}_L i \gamma^\mu\Big(\dmu+
i g {\tilde W}_\mu ^a{{\tau^a}\over {2}}+
		      {{i}\over {2}}\gpt(B-L){\tilde Y}_\mu\Big){\psi}_L\cr
     &+&\overline{\psi}_R i \gamma^\mu\Big(\dmu+
i\gp {\tilde Y}_\mu{{\tau^3}\over {2}}+{{i}\over {2}}\gpt
		      (B-L) \Yt_\mu\Big){ \psi}_R
\eea
where $B(L)$ is the baryon (lepton) number, and 
$\psi =\left(\psi_u,\psi_d\right)$.
We have not introduced direct couplings to  $\Lt$ and $\Rt$, so
the new gauge bosons will couple to fermions only via mixing.

By separating  the charged and the neutral gauge  bosons, 
the quadratic lagrangian is given by:

\bea
{\cal L}^{(2)} &=& {{v^2}\over {4}}[(1+2 a_2)g^2 \tW_\mu^+ \tW^{\mu -}+
		      a_2 \gs^2 (\tL_\mu^+ \tL^{\mu -}+\tR_\mu^+ 
		  \tR^{\mu -})\cr
		&-&\sqrt{2}a_2g \gs (\tW_\mu^+ \tL^{\mu -}+\tW_\mu^- 
		\tL^{\mu +})]\cr
&+& {{v^2}\over {8}}[(1+2 a_2) (g^2 \tW_3^2+\gptd \tY^2)+
		      a_2 \gs^2 (\tL_3^2+\tR_3^2)\cr
		 &-& 2 g \gpt \tW_{3\mu}\tY^\mu
		-2 \sqrt{2}a_2\gs (g \tW_3 \tL_3^\mu +\gp \tY_\mu 
		 \tR_3^\mu)]
\label{2.5}
\eea
Therefore the $R^\pm$ fields are unmixed and their mass can be easily 
read: $M_{R^{\pm}}\equiv M=v \gs  \sqrt  a_2/2$.
All the other  heavy fields have  degenerate mass $M$ in the large
$\gs$ limit (for this reason we call this model Degenerate BESS),
and $W$ and $Z$ masses get corrections of order $(g/\gs)^2$
\cite{dege}.
We will parameterize the model by using, in addition to
the SM parameters, $M$ and $g/\gs$.

By using eq. (\ref{2.5}) one can show that, at the leading 
order in $q^2/M^2$,  the contribution of the model to all
$\eps$ parameters \cite{alta} is equal to zero \cite{dege}.
This is due to the fact that in the $M\to\infty$ limit, this model decouples.
We can perform the low-energy limit at the next-to-leading order and study
the virtual effects of the heavy particles. Working at the first order in 
$1/\gs^2$ we get $\epsilon_1=-(\c^4+\s^4)/(\c^2)~ X$, 
$\epsilon_2=-\c^2~ X$, $\epsilon_3=-X$
with $X=2({M_Z^2}/{M^2})( g/\gs)^2$.
All these deviations are of order $X$ which contains a double
suppression factor $M_Z^2/M^2$ and $(g/\gs)^2$. 
The sum of the SM contributions, functions of the top and Higgs masses,
and of these deviations has to be compared with the experimental
values for the $\epsilon$ parameters, determined from the all 
available LEP data and the $M_W$ measurement at Tevatron \cite{cara}:
$\epsilon_1=(3.8 \pm 1.5)\cdot 10^{-3}$,
$\epsilon_2=(-6.4\pm 4.2)\cdot 10^{-3}$,
$\epsilon_3=(4.6\pm 1.5)\cdot 10^{-3}$.
Taking into account the SM values
$(\epsilon_1)_{SM}=4.4\cdot 10^{-3}$,
$(\epsilon_2)_{SM}=-7.1\cdot 10^{-3}$,
$(\epsilon_3)_{SM}=6.5\cdot 10^{-3}$ for $m_{top}=180~GeV$ and
$m_H=1000~GeV$, we find, from the combinations of the previous 
experimental results, the $90\%$ C.L. limit on $g/\gs$ versus
the mass M given in Fig. 1. The allowed
region is the one below the solid line.

\begin{figure}
\epsfysize=8truecm
\centerline{\epsffile{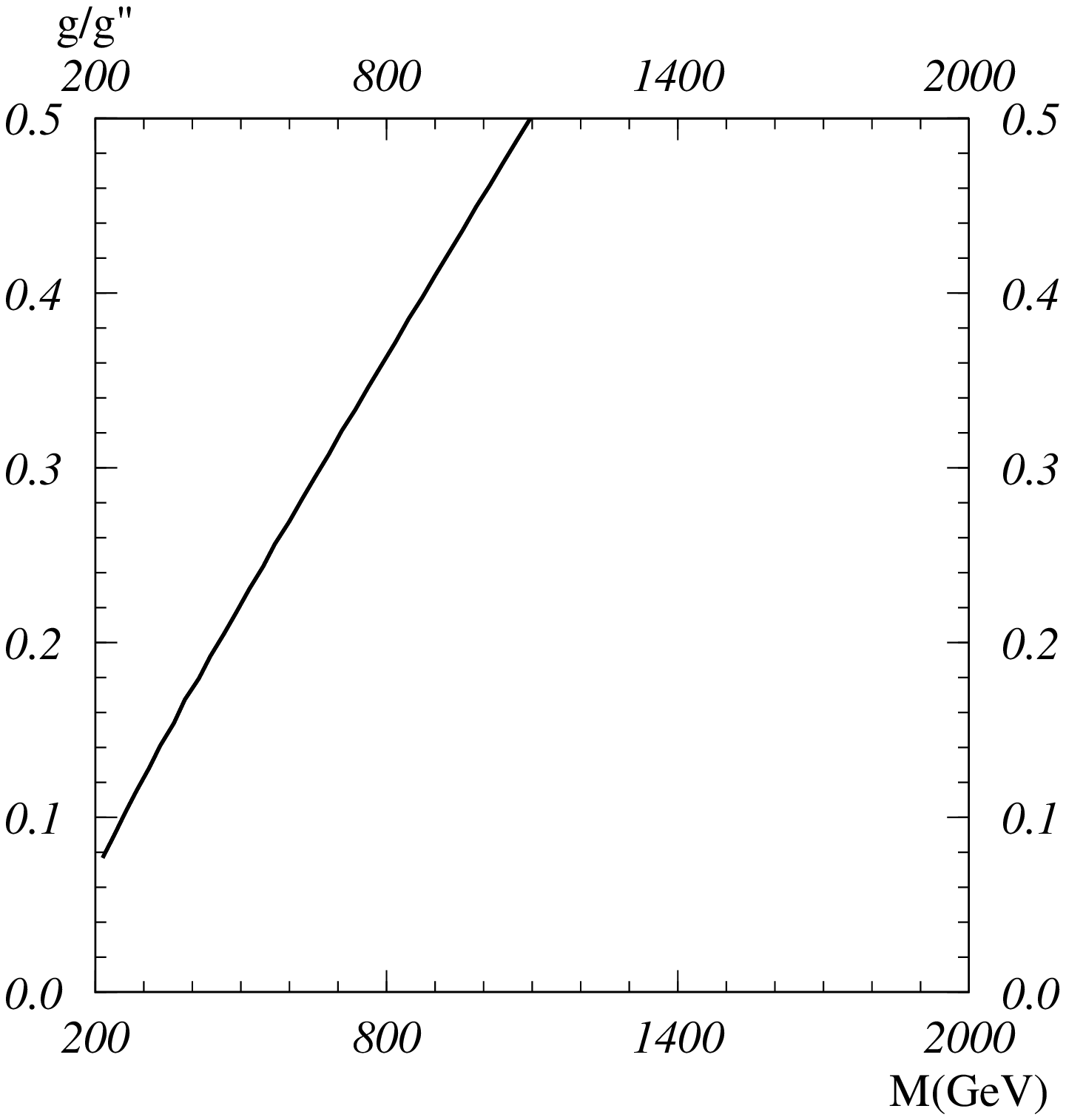}}
\smallskip
\noindent
{\bf Fig. 1} - {\it 90\% C.L. contour on the plane ($M$, $g/g''$) obtained by
comparing the values of the $\epsilon$ parameters from the model
to the experimental data from LEP. The allowed region is below the curve.}
\end{figure}

\section{$e^+e^-$ future colliders}

In this section we will discuss 
the sensitivity of the model at LEP2 and future $e^+e^-$ 
linear colliders, for different options of total centre of mass energies and 
luminosities. 

Cross-sections and asymmetries for the channel 
$e^+e^-\rightarrow f^+f^-$ and $e^+e^-\rightarrow W^+W^-$ in the SM
 and in the degenerate BESS model at tree level have been studied
\cite{dege}. The BESS states relevant for the 
analysis at $e^+e^-$ colliders are $L_3$ and $R_3$. Their coupling
to fermions can be found in \cite{dege}. We will not 
consider the direct production of $R_3$ and $L_3$ from $e^+e^-$,
but rather their indirect effects in the $e^+e^-\rightarrow f^+f^-$ and  
$e^+e^-\rightarrow W^+W^-$ cross-sections.
In the fermion channel the study is based on the following observables:
the total hadronic ($\mu^+\mu^-$) cross-sections $\sigma^h$ ($\sigma^{\mu}$),
the forward-backward and left-right asymmetries  $A_{FB}^{e^+e^- \to \mu^+ 
\mu^-}$, $A_{FB}^{e^+e^- \to {\bar b} b}$, 
$A_{LR}^{e^+e^- \to \mu^+ \mu^-}$, 
$A_{LR}^{e^+e^- \to h}$ and  $A_{LR}^{e^+e^- \to {\bar b} b}$.
At LEP2 we can add to the previous observables the $W$ mass measurement.
The result of this analysis shows that LEP2 will not improve considerably 
the existing limits \cite{lep}. 

\begin{figure}
\epsfysize=8truecm
\centerline{\epsffile{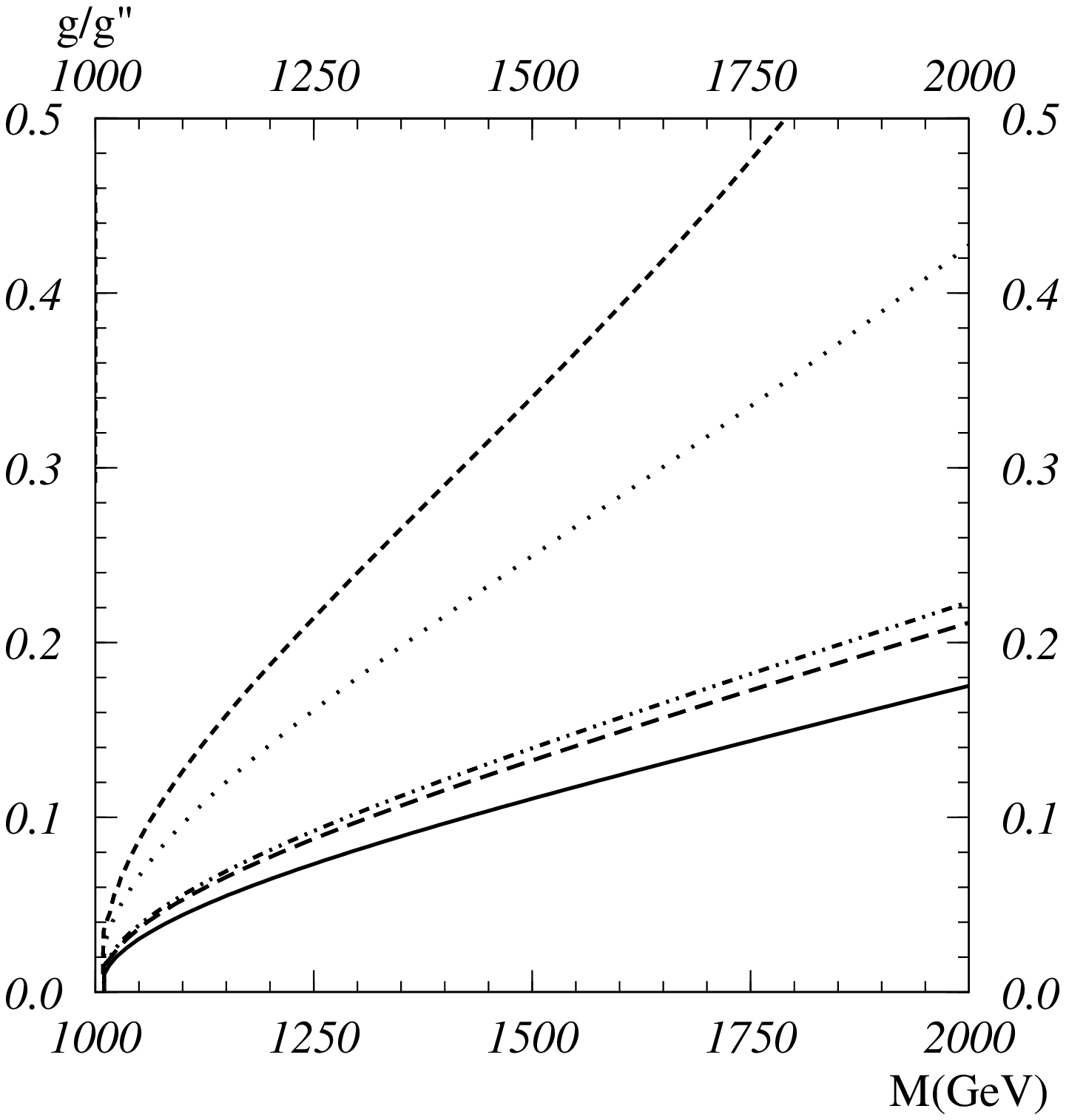}}
\noindent
{\bf Fig. 2} - {\it 
 90\% C.L. contour on the plane 
($M$, $g/g''$) from $e^+e^-$ at $\sqrt{s}=1000~GeV$ with 
an integrated luminosity of $80 fb^{-1}$ from unpolarized observables.
Allowed regions are below the curves.
(Dashed-dotted is $\sigma^h$, dashed is $\sigma^\mu$, dotted is
$A_{FB}^\mu$, the uppermost dashed is $A_{FB}^b$,
continuous is for all combined).}
\end{figure}

In Fig. 2 we present the 90\% C.L. contour on the plane 
($M$, $g/g''$) from $e^+e^-$ at $\sqrt{s}=1000~GeV$ with 
an integrated luminosity of $80 fb^{-1}$ for various 
observables. The dashed-dotted line 
represents the limit from $\sigma^h$ with an assumed relative error of 2\%; 
the dashed line near to the preceeding one is $\sigma^\mu$ (relative 
error 1.3\%), the dotted line 
is $A_{FB}^\mu$  (error 0.5\%) and the uppermost dashed line is $A_{FB}^b$
(error 0.9\%).

As it is evident more stringent bounds come from the 
cross-section measurements. Asymmetries give less restrictive bounds due to
a compensation between the $L_3$ and $R_3$ exchange.
By combining all the deviations in the previously considered
observables we get the limit shown by the continuous line.

\begin{figure}
\epsfysize=8truecm
\centerline{\epsffile{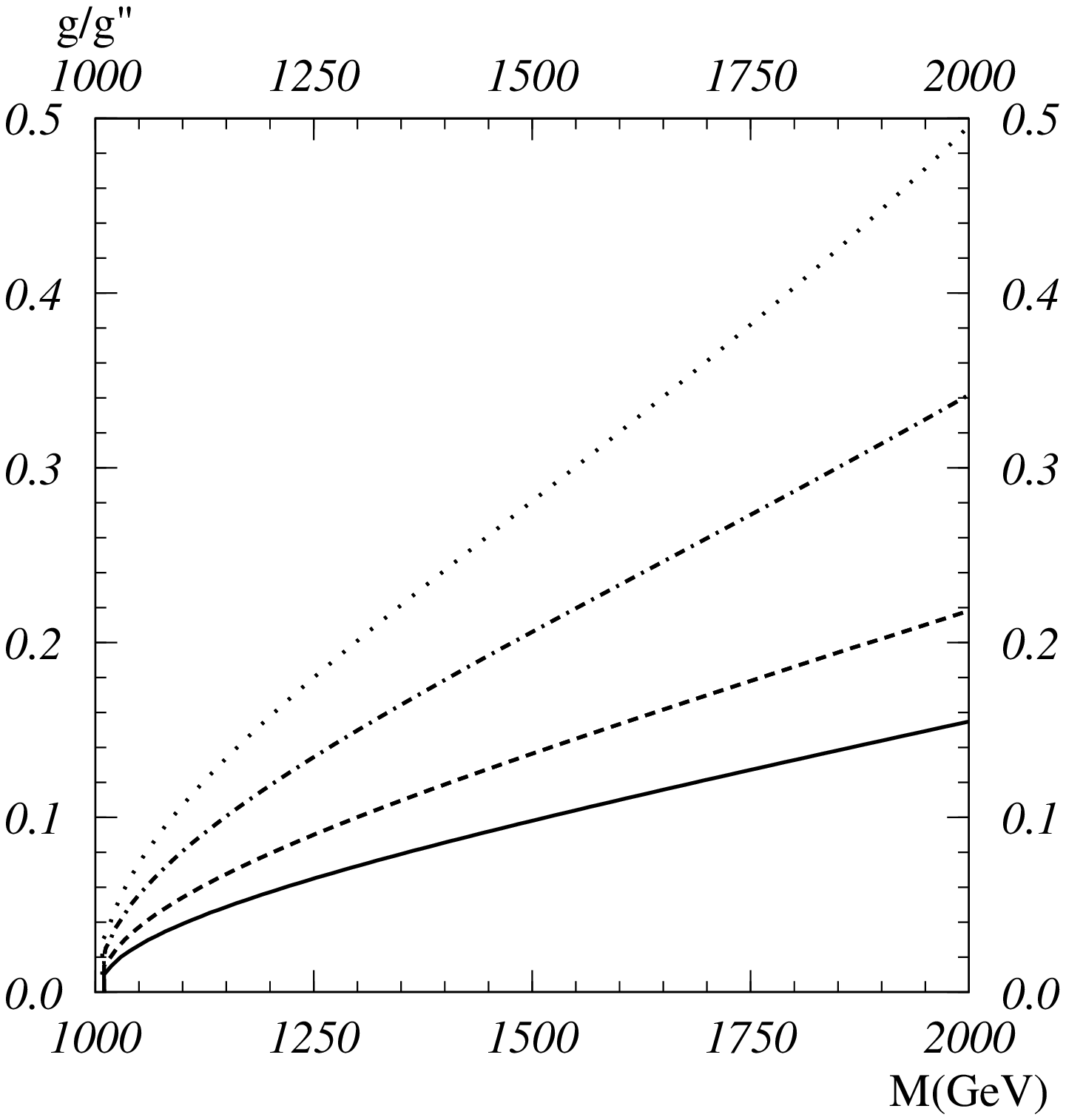}}
\noindent
\noindent
{\bf Fig. 3} - {\it  90\% C.L. contour on the plane 
($M$, $g/g''$) from $e^+e^-$ at $\sqrt{s}=1000~GeV$ with 
an integrated luminosity of $80 fb^{-1}$ from polarized observables.
Allowed regions are below the curves. (Dashed-dotted is $A_{LR}^\mu$, 
dashed is $A_{LR}^h$, dotted is $A_{LR}^b$, continuous is for all 
unpolarized and polarized combined).}
\end{figure}

Polarized electron beams
allow to get further limit in the parameter space as shown in Fig. 3.
We neglect the error on the measurement of the polarization and use a 
polarization value equal to 0.5.
The dashed-dotted line represents the limit from $A_{LR}^\mu$ (error  0.6\%), 
the dashed line from $A_{LR}^h$ 
(error 0.4\%), the dotted line from $A_{LR}^b$ (error 1.1\%).
Combining all the polarized and unpolarized beam observables we get the 
bound shown by the continuous line. In conclusion 
a substantial improvement with respect to the LEP bounds,
even without polarized beams is obtained.

\begin{figure}
\epsfysize=8truecm
\centerline{\epsffile{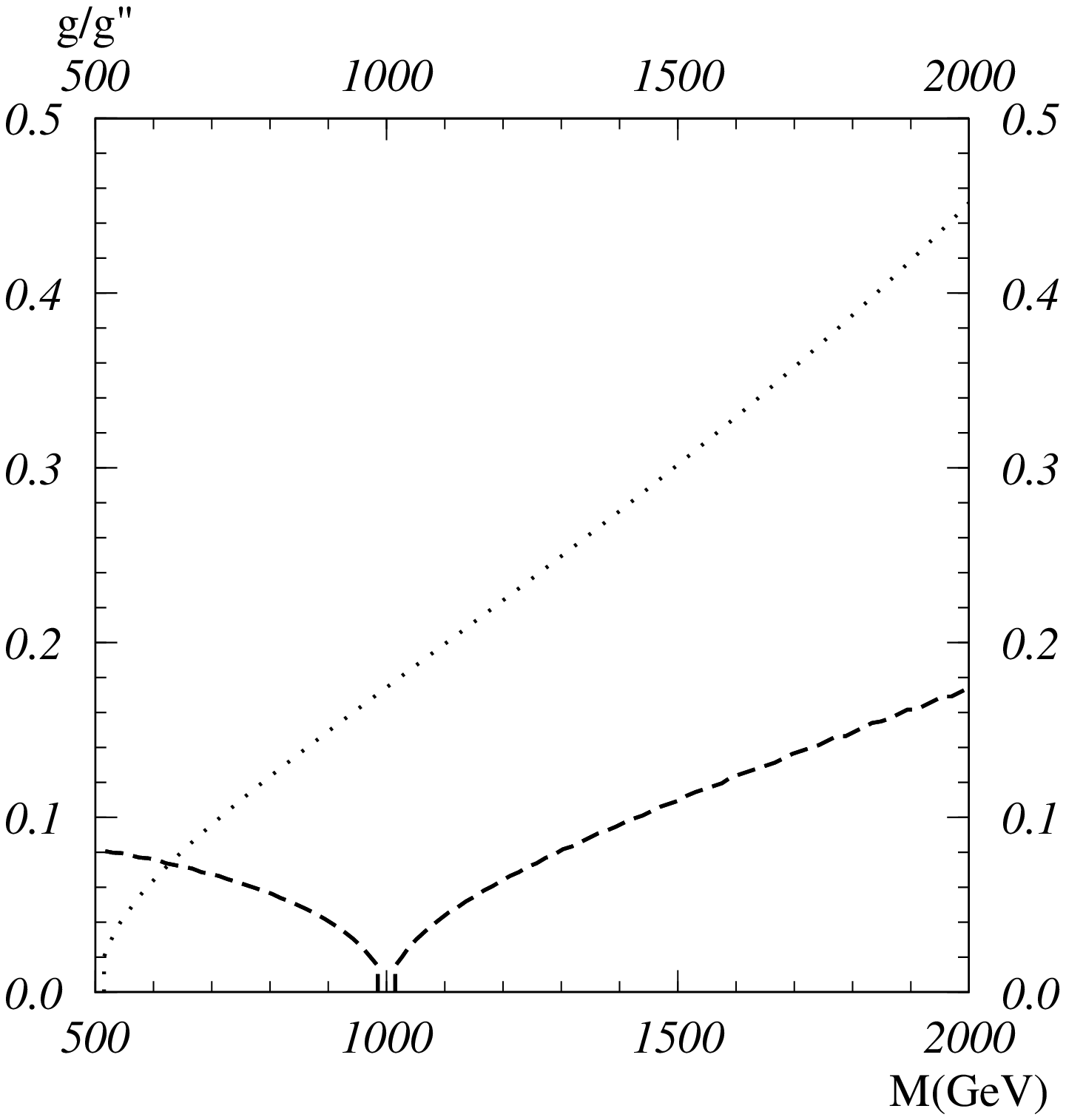}}
\noindent
\noindent
{\bf Fig. 4} - {\it  90\% C.L. contour on the plane ($M$, $g/g''$) from 
$e^+e^-$ at $\sqrt{s}=500~GeV$ with an 
integrated luminosity of $20 fb^{-1}$ and $\sqrt{s}=1000~GeV$ 
with an integrated luminosity of $80 fb^{-1}$.
Allowed regions are below the curves.}
\end{figure}

In Fig. 4 a combined picture of the
90\% C.L. contours on the plane ($M$, $g/g''$) from $e^+e^-$
at two values of $\sqrt{s}$ is shown. The dotted 
line represents the limit from the combined unpolarized observables at 
$\sqrt{s}=500~GeV$ with an integrated luminosity of $20 fb^{-1}$; the
dashed line is the limit from the combined unpolarized observables 
at $\sqrt{s}=1000~GeV$ with an integrated luminosity of $80 fb^{-1}$.
As expected increasing the energy of the collider and rescaling the
integrated luminosity result in stronger bounds on the parameter space. 

The $WW$ final state, considering the observables
given in \cite{dege} has been also studied. However 
the new channel does  not modify the strong limits
obtained using the fermion final state. This is  because
the degenerate model has no strong 
enhancement of the $WW$ channel, present in the usual
strong electroweak models.
For example, this is the most important channel for the BESS model with only 
vector resonances \cite {desy}.

\end{document}